\documentclass[prl,showpacs,superscriptaddress,twocolumn]{revtex4}
\usepackage{dcolumn}
\usepackage{graphicx}

\begin{document}
\raggedbottom

\title{Direct measurement of the radiative lifetime
of vibrationally excited OH radicals}

\author{Sebastiaan Y.T. van de Meerakker}
\affiliation{Fritz-Haber-Institut der Max-Planck-Gesellschaft,
Faradayweg 4-6, 14195 Berlin, Germany} \affiliation{FOM-Institute
for Plasmaphysics Rijnhuizen, Edisonbaan 14, 3439 MN Nieuwegein,
the Netherlands}
\author{Nicolas Vanhaecke}
\affiliation{Fritz-Haber-Institut der Max-Planck-Gesellschaft,
Faradayweg 4-6, 14195 Berlin, Germany}
\author{Mark P.J. van der Loo}
\affiliation{Institute of Theoretical Chemistry, University of
Nijmegen, Toernooiveld 1, 6525 ED Nijmegen, the Netherlands }
\author{Gerrit C. Groenenboom}
\affiliation{Institute of Theoretical Chemistry, University of
Nijmegen, Toernooiveld 1, 6525 ED Nijmegen, the Netherlands }
\author{Gerard Meijer}
\affiliation{Fritz-Haber-Institut der Max-Planck-Gesellschaft,
Faradayweg 4-6, 14195 Berlin, Germany}

\date{\today}

\begin{abstract}
Neutral molecules, isolated in the gas-phase, can be prepared in a
long-lived excited state and stored in a trap. The long
observation time afforded by the trap can then be exploited to
measure the radiative lifetime of this state by monitoring the
temporal decay of the population in the trap. This method is
demonstrated here and used to benchmark the Einstein
$A$-coefficients in the Meinel system of OH. A pulsed beam of
vibrationally excited OH radicals is Stark decelerated and loaded
into an electrostatic quadrupole trap. The radiative lifetime of
the upper $\Lambda$-doublet component of the $X\,^2\Pi_{3/2}, v=1,
J=3/2$ level is determined as $59.0 \pm 2.0$ ms, in good agreement
with the calculated value of $57.7 \pm 1.0$ ms.
\end{abstract}


\maketitle

Infrared absorption and emission spectroscopy has a long and rich
history in the development and application of molecular physics.
Absorption and emission of infrared radiation is an important
diagnostic means in determining the presence and concentration of
molecular species in various environments, ranging from plasmas
and flames to the earth's atmosphere and interstellar space. A
quantitative analysis of these observations relies on a detailed
knowledge of the Einstein $A$-coefficients that characterize the
spontaneous emission rates. These $A$-coefficients can generally
be inferred from absorption measurements, provided the
(line-integrated) number density and the temperature of the
absorbing species are accurately known. The unstable nature of
many chemically highly relevant molecules severely limits the
accuracy of this approach for these species, however. The most
direct and generally applicable route to accurately determining
the $A$-coefficients is to measure the radiative lifetimes of
individual ro-vibrational levels. Molecules isolated in the
gas-phase and at low densities are needed for this, as
interactions with a medium and quenching by collisions have to be
avoided. The preferred way to measure the radiative lifetime is to
prepare molecules in the quantum state of interest, and to measure
the state-specific population as a function of time. The problem
with this approach is that infrared radiative lifetimes are
typically in the millisecond to second range, much longer than the
observation times that are commonly available in experiments.
Ingenious schemes have been developed that nevertheless enable a
measurement of long radiative lifetimes of neutral molecules in a
molecular beam \cite{Drabbels1997,Jongma1997/1}. Only for
lifetimes up to a few milliseconds it has been possible to measure
the population decay directly \cite{Jongma1997/2}.

For charged particles, the greatly enhanced observation times that
became available when traps were developed opened up the
possibility to directly measure radiative lifetimes of metastable
states \cite{Schneider1979}. For neutral atoms and neutrons,
trapping has been used to directly measure long lifetimes as well
\cite{Huffman2000}. Methods to confine neutral gas-phase molecules
in magnetic \cite{Weinstein1998}, optical \cite{Takekoshi1998},
and electrostatic traps \cite{Bethlem2000} for times up to seconds
have recently been developed. In this Letter we report on the
first direct measurement of the infrared radiative lifetime of a
vibrationally excited trapped molecule. By measuring the temporal
decay of the population of OH ($X\,^2\Pi_{3/2}, v=1, J=3/2$)
radicals in an electrostatic trap, an accurate value for the
$A$-coefficient of the fundamental $1-0$ band of OH is obtained.

The infrared radiative properties of the OH radical are of
particular importance. Vibrationally excited OH ($X\,^2\Pi, v\leq
9$) radicals, produced in the upper atmosphere via the reactive
depletion of ozone \cite{Summers1997,Wennberg1994}, are
responsible for the near-infrared night-time air-glow
\cite{Meinel1950}. Recently, this OH Meinel band emission has also
been observed from artificial aurora at higher altitudes, offering
new possibilities to study ionospheric interactions
\cite{Kagan2005}. OH vibrational emission has also been observed
from stellar and interstellar space \cite{Wilson1968}. The
so-called `prompt emission' in the $1-0$ band of the OH radical
(around 3.3~$\mu$m), produced by photodissociation of water, is
used as a tracer for water produced in comets \cite{Bonev2004}.
The radiative lifetimes of the vibrational states of OH
($X\,^2\Pi$) are essential for a quantitative interpretation of
all these observations, and have therefore received considerable
experimental and theoretical attention. Over the years, the values
have scattered over a wide range and have only slowly converged
\cite{Nelson1990}. The most recent values for the lifetimes, based
on experimental absorption line intensities, are given in the
\texttt{HITRAN 2004} database \cite{rothman:05}. For the
$X\,^2\Pi_{3/2}, v=1, J=3/2$ level of OH, this database gives a
lifetime of $56.6$ ms with an error of 10-20\%.

\begin{figure}
    \centering
    \resizebox{\linewidth}{!}
    {\includegraphics[0,0][524,346]{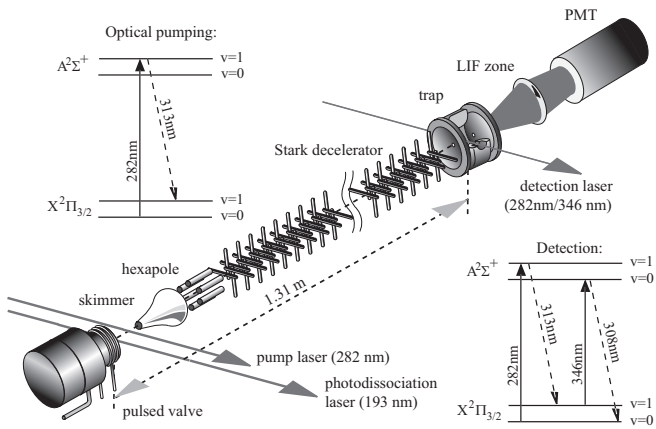}}
    \caption{Scheme of the experimental setup. A pulsed beam of
    OH ($v=0$) radicals is produced via photodissociation (193~nm) of HNO$_3$
    molecules seeded in Xe. Prior to entering the decelerator, part of the OH radicals are
    prepared in the $v=1$ state by Franck-Condon pumping via the
    $A\,^2\Sigma^+$ state (upper left). OH radicals in the $J=3/2$ level of
    both the $v=0$ and $v=1$ vibrational states of the $X\,^2\Pi_{3/2}$
    electronic ground state are Stark
    decelerated and subsequently loaded into the electrostatic
    trap. Molecules in the trap are state-selectively detected
    by imaging the laser induced fluorescence onto a PMT.}
    \label{fig:Setup}
\end{figure}

The experiments are performed in a molecular beam deceleration and
trapping machine that is schematically shown in Figure
\ref{fig:Setup}.
The same setup has been used recently for the experimental
demonstration of the electrostatic trapping of OH
($X\,^2\Pi_{3/2}, v=0$) radicals \cite{Meerakker2005}. A pulsed
beam of OH radicals with a mean velocity of 360~m/s is produced
via ArF-laser dissociation of HNO$_3$, seeded in Xe, near the
orifice of a pulsed solenoid valve. In the experiments, the
population in the $J=3/2$ rotational level of the $v=0$ and $v=1$
vibrational states of the $X\,^2\Pi_{3/2}$ electronic ground state
is of relevance. During the expansion, most of the OH radicals
cool down to the ro-vibrational ground state $v=0, J=3/2$. Just
after the expansion region, approximately 15\% of the OH ($v=0$)
radicals are promoted to the $X\, ^2\Pi_{3/2}, v=1$ state by
Franck-Condon pumping via the $A\,^2\Sigma^+$ state at 282~nm
\cite{Jaffer1979}. This optical pumping scheme is shown in the
upper left of Figure \ref{fig:Setup}. In the presence of an
electric field, the upper $\Lambda$-doublet components of the
$J=3/2$ levels split into a low-field seeking $M_J\Omega=-3/4$ and
a $M_J\Omega=-9/4$ component. Only OH radicals in the $J=3/2,
M_J\Omega=-9/4$ component of either vibrational state are Stark
decelerated and subsequently loaded into the quadrupole trap.

After passage through a 2~mm diameter skimmer, the beam enters the
deceleration chamber, and is focussed into the Stark decelerator
by a short hexapole. The operation principles of the hexapole,
Stark decelerator, and quadrupole trap are given elsewhere
\cite{Meerakker2005}. Briefly, the Stark decelerator exploits the
interaction of polar molecules with time-varying electric fields
to reduce the velocity of the molecular beam in a stepwise
fashion. The 1.18~m long decelerator consists of an array of
108~electric field stages, that are placed 11~mm apart. Each field
stage consists of two 6~mm diameter parallel electrodes, centered
10~mm apart that are oriented perpendicular to the molecular beam
axis. Adjacent electric field stages are rotated by 90$^{\circ}$
with respect to each other to provide a $4\times 4$ mm$^2$
transverse area for the beam to pass. When at the appropriate
times a voltage difference of 40~kV is applied between the
electrodes of the electric field stages, OH radicals in the
$J=3/2, M_J\Omega=-9/4$ component are decelerated from 360~m/s to
20~m/s. The Stark shifts of the $v=0$ and $v=1$ states of OH are
very similar \cite{Peterson1984}, and molecules in both
vibrational states are simultaneously decelerated and loaded into
the quadrupole trap with equal efficiency. The trap consists of a
ring electrode and two hyperbolic end-caps, and is located 21~mm
downstream from the decelerator. The trap is mounted in a separate
differentially pumped vacuum chamber, where the pressure does not
exceed $2 \times 10^{-9}$ mbar under operating conditions.
Depending on the voltages that are applied to the trap electrodes,
either a last deceleration stage or a (nearly) symmetric 500~mK
deep potential well can be created. The OH radicals enter the trap
through a 4~mm diameter hole in the first end-cap, and come,
approximately 8~ms after their production, to a standstill near
the center of the trap. At this time, the voltages on the trap
electrodes are switched to confine the OH radicals in the
potential well. The density of OH radicals at the center of the
trap is measured using laser induced fluorescence (LIF) detection
schemes, shown in the lower right of Figure \ref{fig:Setup}. The
population of the OH radicals in the $v=0, J=3/2$ and $v=1, J=3/2$
levels is state-selectively probed by inducing the $Q_1(1)$
transitions of the $A\,^2\Sigma^+, v=1 \leftarrow X\,^2\Pi, v=0$
band around 282~nm and the $A\,^2\Sigma^+, v=0 \leftarrow
X\,^2\Pi, v=1$ band around 346~nm, respectively. The off-resonant
fluorescence around 313~nm (when probing $v=0$) and 308~nm (when
probing $v=1$) is imaged through a 6~mm diameter opening in the
second end-cap onto a photomultiplier tube (PMT). Stray light from
the laser is largely avoided by using light baffles and is
suppressed by optical filters in front of the PMT.

\begin{figure}
    \centering
    \resizebox{\linewidth}{!}
    {\includegraphics[0,0][405,305]{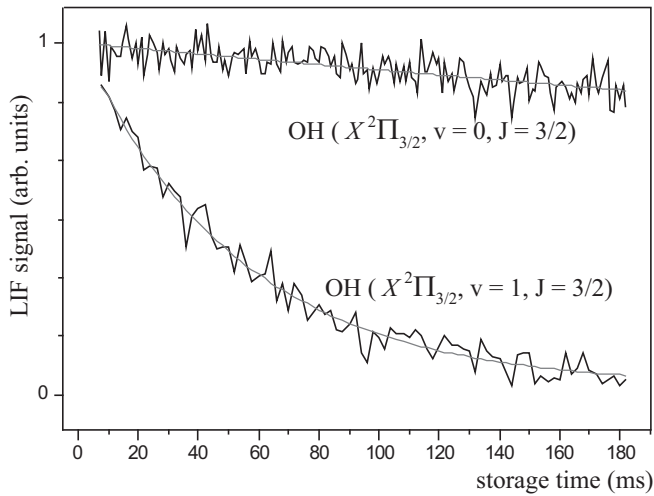}}
    \caption{Population of trapped OH radicals in the
    $X^2\Pi_{3/2},v=0,J=3/2$ level (upper curve) and in the
    $X^2\Pi_{3/2},v=1,J=3/2$ level (lower curve), as a
    function of the storage time $t$. The trapping potential
    is switched on at $t=0$ ms.}
    \label{fig:storage-time}
\end{figure}

In Figure \ref{fig:storage-time}, the LIF signal of the trapped
cloud of OH radicals in the $v=0, J=3/2$ level (upper curve) and
in the $v=1, J=3/2$ level (lower curve) is shown as a function of
the storage time $t$. The trapping potential is switched on at
$t=0$ ms. The experiment runs at a repetition rate of 5~Hz to
allow for a maximum observation time of (almost) 200~ms in the
trap.
On this time-scale, OH radicals in the $v=0$ state mainly leave
the trap via collisions with background gas; both elastic and
inelastic collisions with the thermal background gas transfer an
amount of kinetic energy to the OH radicals that largely exceeds
the trap depth. From a separate series of measurements, running
the apparatus at lower repetition rates while maintaining the same
background pressure, a 1/$e$ trap lifetime of $1.3 \pm 0.1$ s is
deduced. The population of trapped OH radicals in the $v=1$ state,
on the other hand, is mainly depleted via spontaneous emission to
the $v=0$ state, in addition to collisions with background gas.
The observed exponentially decaying curve therefore almost
directly reflects the radiative lifetime of the $X\,^2\Pi_{3/2},
v=1, J=3/2$ level of OH. From a series of measured decay curves,
taking carefully the baseline and the signal-to-noise ratio of the
LIF signal into account, an exponential decay time of $56.4 \pm
1.9$ ms is deduced. The inherent trap lifetime can be assumed to
be identical for OH radicals in the $v=0$ and $v=1$ vibrational
states, as the total collision cross-sections with thermal
background gas will be very similar for OH in either one of the
vibrational states. Under this assumption, a radiative lifetime
$\tau$ for OH radicals in the upper $\Lambda$-doublet component of
the $X\,^2\Pi_{3/2}, v=1, J=3/2$ level of $\tau=59.0 \pm 2.0$ ms
is obtained. We verified that the re-population of the $v=1$ state
in the trap via cascading spontaneous emission from higher
vibrational states is negligible in our experiment.

OH radicals in the $v=1, J=3/2$ level that are confined in the
trap undergo infrared spontaneous emission to the $v=0$ state,
following the selection rules of the electric dipole allowed
transitions in the presence of an electric field. The molecules
can make a transition to all $M_J\Omega$ components of the
$J=1/2$, $J=3/2$ and $J=5/2$ rotational levels of the $X\,^2\Pi,
v=0$ electronic ground state. Of these, only molecules that end up
in the $M_J\Omega=-15/4$, $-9/4$ and $-3/4$ components of the
$J=5/2$ level or in the $M_J\Omega=-9/4$ and $-3/4$ component of
the $J=3/2$ level are (partially) recaptured and confined in the
trap.
\begin{figure}
     \centering
     \resizebox{\linewidth}{!}
     {\includegraphics[0,0][414,340]{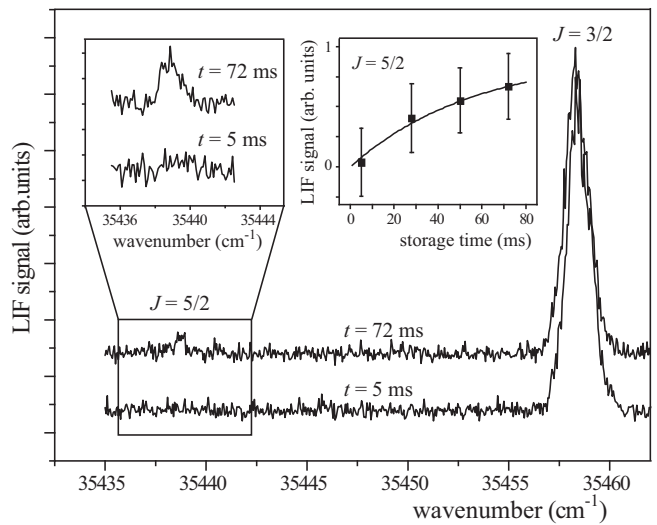}}
     \caption{Excitation spectrum of trapped OH ($v=0$) radicals
     that is recorded $5$~ms (lower curve) and $72$~ms
     (upper curve) after switching on of the trap. In the central inset,
     the population of trapped OH radicals in the $v=0,J=5/2$ level is shown
     as a function of the storage time.}
     \label{fig:increase52}
\end{figure}
In Figure \ref{fig:increase52}, the excitation spectrum of trapped
OH ($v=0$) radicals is shown that is recorded $5$~ms (lower curve)
and $72$~ms (upper curve) after switching on of the trap. The
detection laser is scanned over the $Q_1(1)$ and $Q_1(2)$
transitions of the $A^2\Sigma^+,v=1\,\leftarrow \, X^2\Pi_{3/2},
v=0$ band, probing the population in the $v=0,J=3/2$ and
$v=0,J=5/2$ levels, respectively. Initially, the trapped OH $v=0$
radicals exclusively reside in the $J=3/2$ level, the only
rotational state that is selected by the Stark decelerator. After
72~ms of trapping, a small but significant population of OH
radicals in the $v=0,J=5/2$ level is detected in the trap. In the
central inset of Figure \ref{fig:increase52}, the accumulation
over time of the trapped OH ($v=0,J=5/2$) radicals is shown. The
expected exponential growth of the population in the $J=5/2$
level, as deduced from the measurements shown in Figure
\ref{fig:storage-time}, is shown as well. The relative population
of trapped OH radicals in the $v=0, J=3/2$ and $v=0, J=5/2$ levels
is also consistent with the expectations. An overall factor of
about 20 is expected, based on the initial relative population of
$v=0$ and $v=1$ in the trap and the H\"{o}nl-London factor for
fluorescence from the $v=1, J=3/2$ to the $v=0, J=5/2$ level, in
combination with the larger size of the trapped cloud of $J=5/2$,
e.g. the reduced overlap with the detection laser.

The radiative properties of the OH radical can be calculated from
its dipole moment function and the potential energy curve. The
radiative lifetime $\tau$ of the upper $\Lambda$-doublet component
(of $f$(+) parity) of the $X^2\Pi_{3/2}(v=1, J=3/2)$ level of OH
is given by
\begin{equation}
\label{eq:tau}
  \tau^{-1} = \sum_f A^{fi} = \sum_f \frac{4\alpha\omega_{fi}^3}{3c^2e^2} |\langle f|\mu|i\rangle|^2,
\end{equation}
where $A^{fi}$ are the Einstein $A$-coefficients for the
$X\,^2\Pi, v=1 \rightarrow X\,^2\Pi, v=0$ transitions, $\alpha$ is
the fine structure constant, $\omega_{fi}$ the frequency of the
emitted photon, $c$ the speed of light, and $e$ the elementary
charge. In the calculation of the initial ($|i\rangle$) and final
($|f\rangle$) states, $\Lambda$-type doubling and spin-orbit,
spin-rotation, and rotation-vibration coupling is taken into
account. The sum over final states includes all $F_1$ and $F_2$
states with $J\le 5/2$. The $X^2\Pi$ potential energy curve and
the $r$-dependent dipole moment $\mu$ are obtained from an
internally contracted multireference configuration interaction
calculation with the \texttt{MOLPRO} \cite{molpro:98} program
package. Special relativistic effects are taken into account by
using Douglas-Kroll one-electron integrals. The molecular orbitals
are obtained from a 1-5$\sigma$, 1-2$\pi$ complete active space
self consistent field calculation, employing an augmented
correlation consistent polarized valence 6-zeta (aug-cc-pV6Z)
one-electron basis set. The $r$-dependent spin orbit couplings are
obtained in a similar calculation employing an uncontracted
aug-cc-pVQZ basis. The $\Lambda$-type doubling parameters are
taken from spectroscopic observations \cite{melen:95}. The result
that is obtained for the radiative lifetime is $\tau=57.7$ ms, in
good agreement with the experimentally determined value. As the
computed ro-vibrational energy levels agree with experiments to
better than 0.02\% and as our dipole moments agree with previous
high level {\em ab initio} calculations \cite{langhoff:89} to
about 0.7\%, we estimate the error in $\tau$ to be about 1~ms.

In this Letter, we report the first direct measurement of the
quantum-state specific infrared radiative lifetime of a trapped
molecule. The technique reported here can generally be applied to
directly measure lifetimes of metastable states of neutral
molecules up to seconds, with an unprecedented accuracy. A
requirement is that it must be possible to confine the molecules
in the desired quantum state in a trap. The molecules can be
prepared in the metastable state prior to deceleration and
trapping, as reported here. Alternatively, they can first be
trapped in the quantum-state that is best suited for it, and
preparation of the desired metastable state can then be performed
inside the trap. By recording the decay of the population of
vibrationally excited OH ($X\,^2\Pi_{3/2}, v=1, J=3/2$) radicals
in an electrostatic quadrupole trap, the radiative lifetime of
this level is determined as $59.0\pm 2.0$~ms, in good agreement
with the theoretically predicted value of $57.7\pm 1.0$~ms. This
provides a benchmark for the Einstein $A$-coefficients in the
Meinel system of OH. This is of particular importance in, for
instance, combustion, atmospheric science and in astrophysics.

\begin{acknowledgments}
This work is part of the research program of the `Stichting voor
Fundamenteel Onderzoek der Materie (FOM)' and of `Chemische
Wetenschappen (CW)', which are financially supported by the
`Nederlandse Organisatie voor Wetenschappelijk Onderzoek (NWO)'.
This work is supported by the EU "Cold Molecules" network. We
acknowledge the assistance of J. Gilijamse and B. Sartakov for
this work.
\end{acknowledgments}

\end{document}